\documentstyle[aps,prl,multicol,epsf,epsfig,float]{revtex}

\newcommand{\ra}{{\rm a}}

\newcommand{\hc}{H_{\rm cond}}

\title{Quantum Computing Using Dissipation to Remain in a Decoherence-Free Subspace}
\author{Almut Beige$^{(1)}$, Daniel Braun$^{(2)}$, Ben Tregenna$^{(1)}$ and
Peter L. Knight$^{(1)}$}
\address{$(1)$ Optics Section, Blackett Laboratory, Imperial College
London,
London SW7 2BZ, England. \\
$(2)$ FB7, Universit\"at-GHS Essen, 45117 Essen, Germany.}

\begin{document}

\maketitle
\draft

\begin{abstract}
\begin{center}
\parbox{14cm}{We propose a new approach to the implementation of quantum
gates in which  decoherence during the gate operations is strongly
reduced. This is achieved by making use of an environment induced quantum Zeno
effect that confines the dynamics effectively to a decoherence-free
subspace.}
\end{center}
\end{abstract}

\vspace*{0.2cm}
\noindent

\begin{multicols}{2}

Quantum computing has attracted much interest since it became clear that
quantum computers are in principle able to solve hard computational
problems
more efficiently than present classical computers
\cite{Deutsch85,Shor94,Grover97}. The main obstacle inhibiting realizations
arises from the difficulty of isolating a quantum  mechanical
system from its environment. This leads to decoherence and the loss of
information stored in the system, which limits for instance factoring to
small numbers \cite{PleKni}. Schemes have been proposed to correct
for errors induced by decoherence and other imperfections
\cite{Steane96}. Alternatively, the use of decoherence-free subspaces
\cite{Palma,Zanardi97,Lidar98,Guo98} has been proposed for which the
dependence on error correction codes may be much reduced. Nevertheless, the
error rate of each operation must not exceed $10^{-5}$ if quantum computers
are ever to work fault-tolerantly \cite{PresShor}.

In contrast to the widely held folk belief that decoherence is to be
avoided, we show here that dissipation can be used to implement
nearly {\em decoherence-free} quantum gates with a success rate which
can, at least in principle, be arbitrarily close to unity. The main
requirement for this to work is the existence of a decoherence-free
subspace (DFS) in the system under consideration. States in the DFS will be called decoherence-free (DF) states. Examples of DFS are known \cite{Guo98,Zanardi99}, but until now, it was not known how to manipulate states {\em within} a DFS in general \cite{manip}.

In this Letter we propose a concrete example of a DFS whose
states can be used to obtain DF qubits for quantum computing.
In contrast to earlier proposals, we assume that all other states couple
{\em strongly} to the environment. A state with no overlap with DF states
should (nearly immediately) lead to dissipation. We show
that we can interpret the effect of the environment on the system as
that of rapidly repeated measurements of whether the system is DF or
not. This effect, which we call an
{\em environment induced quantum Zeno effect} \cite{misra}, leads to the
fact that a weak interaction only changes the state of the system
{\em inside} the DFS. This allows for a wide range of new
possibilities to perform DF gate operations between the qubits. As an
example we describe a CNOT operation between two qubits
that is almost DF yet rather simple: A {\em single}
laser pulse suffices. We will show that the system proposed fulfills all
criteria for a quantum computer proposed by DiVincenzo \cite{Vinc}.

The system we propose consists of $N$ identical three-level atoms with
a $\Lambda$ configuration. We denote the split ground states of atom $i$ by
$|0\rangle_i$ and $|1\rangle_i$, and the excited state by $|2
\rangle_i$. The
atoms are assumed to be stored in a line, which can be for instance in a
linear ion trap, an optical lattice or on top of a wire on an atom chip
\cite{Folman99}. To realise a gate operation between two
neighbouring atoms (denoted by $i=1$ and $i=2$ in the
following), requires to move them into a cavity, as shown in
Fig.~\ref{fig.sys}. This can be done by moving the lattice or by
applying an electric field, respectively.
We assume that only the atomic 1-2
transition is in resonance with a single resonator mode. For simplicity
the coupling constants of both atoms to the cavity field mode
is taken to be the same, $g_1 = g_2 \equiv g$, but this is not crucial to
our analysis.

\noindent
\begin{minipage}{3.38truein}
\begin{center}
\begin{figure}[h]
\epsfig{file=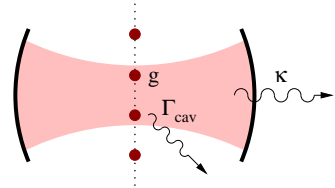,width=4.3cm} \\[0.2cm]
\caption{Schematic view of the system. To perform a gate operation two
three-level atoms are moved to fixed positions inside a
cavity.}\label{fig.sys}
\end{figure}
\end{center}
\end{minipage}
\vspace*{0.1cm}

The environment consists of a continuum of electromagnetic
field modes
surrounding the atoms and the cavity. This gives rise to decoherence in two
different ways: First, individual spontaneous emission of the atoms outside
the cavity can take place with a rate $\Gamma$. For atoms inside
the cavity this rate can be decreased to below its free-space value and
will be denoted by $\Gamma_{\rm cav}$. In addition, the resonant
field mode inside the cavity couples to the outside, given non-ideal
mirrors. A photon
inside the resonator leaks out through the cavity mirrors with a rate
$\kappa$.

To describe the time evolution of the system and to find a simple criterion
for a DFS we will make use of a quantum jump description \cite{HeWi}. This
method gives the time evolution under the condition that no photon is
emitted, as well as the probability for no photon emission, $P_0(t,\psi)$,
where $|\psi\rangle$ is the state of the system at time $t=0$. The system
dynamics is described by a
non-Hermitian Hamiltonian $H_{\rm cond}$ that incorporates the coupling to
the environment. It can be derived rigorously from the full Hamiltonian. Due to the non-Hermiticity, the norm of the
state vector
\begin{eqnarray} \label{phit}
|\psi^0 (t) \rangle
&=& {\rm e}^{- {\rm i} H_{\rm cond} t/\hbar} |\psi \rangle
\end{eqnarray}
decreases with time. The probability $P_0$ to observe {\em no} photon up to
time $t$ by a broadband detector of unit efficiency is given by the squared
norm
\begin{equation} \label{29}
P_0 (t,  \psi) = \langle \psi^0(t)|\psi^0(t) \rangle ~.
\end{equation}
The negative derivative of $P_0$ at time $t=0$ gives the probability
density for an immediate photon emission from state $|\psi
\rangle$ and equals
\begin{eqnarray} \label{I}
I(\psi) &=& {{\rm i} \over \hbar} \, \langle \psi| \, H_{\rm cond}
- H_{\rm cond}^\dagger \, |\psi \rangle ~.
\end{eqnarray}
If no photon is observed, the state of the system at time $t$ is the
state (\ref{phit}) normalised to unity.

In the following $b$ denotes the annihilation operator for
one photon in the cavity mode. If we choose the interaction picture in a
way that the atoms and the cavity mode plus environment are considered as
the free system one finds in a similar way as in
Ref.~\cite{alt} that the conditional Hamiltonian equals
\begin{eqnarray} \label{Hcond}
H_{\rm cond}&=& {\rm i} \hbar \, g
\sum_{i=1}^2 \Big[ \, b \, |2 \rangle_{ii}\langle 1| - {\rm h.c.} \, \Big]
- {\rm i} \hbar \, \Gamma_{\rm cav} \sum_{i=1}^2 |2\rangle_{ii}\langle 2|
\nonumber \\
& & - {\rm i} \hbar \, \Gamma \, \sum_{i=3}^N |2\rangle_{ii}\langle 2|
- {\rm i} \hbar \, \kappa \, b^\dagger b ~.
\end{eqnarray}

According to the above, a simple criterion for a DF state of the atoms and
the cavity field mode is: No photon should be emitted, either by
spontaneous emission or by leakage of a photon through the cavity
mirrors. A state $|\psi \rangle$ belongs to
the DFS iff
\begin{equation} \label{crit}
P_0 (t, \psi) \equiv 1 ~~ \forall ~~ t \ge 0 ~.
\end{equation}
As can be seen from Eq.~(\ref{I}) no
photon emission is possible if the atoms are all in a ground state and the
cavity field is empty. The interaction of the system with the environment
is effectively switched off \cite{Lidar98}. In addition, there is no
energy in the system which can be emitted in form of a photon. The
$2^N$ DF ground states of the system are therefore ideally suited as the $N$
DF qubit memory for the quantum computer \cite{Pellizzari95}. The $i$th
qubit is formed by the two
ground states $|0\rangle_i$ and $|1\rangle_i$ of atom $i$ while there is
no photon in the resonator mode.

In addition to these states we obtain more DF states if we neglect
spontaneous emission by the atoms inside the cavity. These states should
only become populated during gate operations
and allow for nearly DF gates. For $\Gamma_{\rm cav}=0$ given
Eq.~(\ref{I}) no photon emission can also take place if a state
$|2\rangle_i$ $(i=1,2)$ of the atoms
inside the cavity is excited. However, this is not yet sufficient. The
cavity mode must {\em never} become populated, i.e.~the system's own time
evolution must not drive states out of the DFS \cite{Lidar98}.
In the following we denote by $|{\rm n} \varphi \rangle\equiv |{\rm
n}\rangle\otimes |\varphi\rangle $ a state with
$n$ photons in the cavity and the atoms in state $|\varphi \rangle$. A
state $|0 \varphi\rangle$ is
DF if all matrix elements of the form $\langle {\rm n} \varphi' |
\,\hc\, |0 \varphi \rangle$ vanish for $n \ne 0$ and arbitrary
$\varphi'$. This is the case iff
\begin{equation} \label{DFSphi}
J_-\, |\varphi\rangle\equiv \sum_{i=1,2} |1 \rangle_{ii}\langle 2|
\varphi\rangle = 0 ~.
\end{equation}
Besides the superpositions of the atomic ground states the atoms inside
the cavity can also be in a superposition with the {\em trapped} state
$|\ra\rangle \equiv \left( |1\rangle_1 |2\rangle_2
- |2\rangle_1 |1\rangle_2 \right)/\sqrt{2}$,
a {\em maximally entangled} state of the two atoms \cite{alt,Rad}.

For $\Gamma_{\rm cav}=0$ one finds from Eq.~(\ref{Hcond}) and
(\ref{DFSphi})
that $H_{\rm cond} \, |\psi \rangle =0$. Without an additional interaction
a DF state does not change in time. To manipulate the states inside the
DFS
a {\em weak} interaction can be used.
But before we discuss the effect of this interaction
we need to study the effect of the environment on the system in more
detail. Let us define the time $\Delta
T$ as the minimum time in which a system in an arbitrary state outside the
DFS definitely emits 
a photon. Then we can interpret the
observation of the free radiation field outside the cavity over a time
interval $\Delta T$ as a measurement of whether the system is in a DF
state or not. The outcome of the measurement is indicated by the emission
of a photon (no DF state), or its absence (DF state).

Here, the cavity field interacts continuously with its environment and the
system behaves like a system under continuous observation,
e.g.~the time between two consecutive measurements is zero. In such a case
the quantum Zeno effect \cite{misra} can be used to predict the time
evolution of the system in the presence of a {\em weak} interaction which
tries to change the state of the system. The quantum Zeno effect is a
consequence of the projection postulate for ideal measurements and
suggests that any process that would lead out of the DFS is ``frozen'' by
the measurements, which always project the system back into a DF state. In
this way the interaction with the environment protects the system against
dissipation. On the other hand, the dynamics {\em within} the DFS is
insensitive to the measurements and takes place almost unmodified.

In the following $H_{\rm cond}$ describes the conditional
time evolution of the system in the presence of an interaction. As long as
the interaction is weak enough, the effect of the environment on the system
can still be interpreted to a good approximation as rapidly repeated
measurements. Therefore the time development operator over the small time
interval $\Delta T$ is given by $I\!\!P_{\rm DFS} \, U_{\rm
cond}(\Delta T,0) \, I\!\!P_{\rm DFS}$, where $I\!\!P_{\rm DFS}$ is the projector on the decoherence-free subspace. This leads to the {\em effective}
Hamiltonian
\begin{equation} \label{Heff}
H_{\rm eff}=I\!\!P_{\rm DFS} \, H_{\rm cond}
\, I\!\!P_{\rm DFS} ~,
\end{equation}
which has a very different effect compared to atoms in free space.

As an example and to show how to realise a CNOT gate we consider a weak
laser pulse applied to the atoms inside the cavity only. The atoms should
be spatially well separated so that the laser pulses can be applied to
each atom individually. The complex Rabi frequencies for the $j$-2
transition $(j=0,1)$ of atom $i$ $(i=1,2)$ are denoted by
$\Omega_j^{(i)}$. If a laser irradiates the atoms the Hamiltonian
\begin{equation} \label{Hlaser}
H_{\rm laser\,I}={\hbar \over 2} \sum_{i=1}^2 \sum_{j=0}^1 \,
\Big[ \, \Omega_j^{(i)} \, |j \rangle_{ii}\langle 2|
+ {\rm h.c.} \, \Big]
\end{equation}
has to be added to the right hand side of Eq.~(\ref{Hcond}).
The Rabi frequencies $\Omega_j^{(i)}$ set the time
scale on which the state of the atoms are changed due to the laser. This
time must be much longer than the measurement time $\Delta T$
which is of the order of $1/\kappa$ and $\kappa/g^2$.
In addition spontaneous emission by the atoms has to be
negligible during the gate operation which leads to the condition
\begin{equation}
\Gamma_{\rm cav} \ll |\Omega_j^{(i)}| \ll \kappa ~~{\rm and} ~~ g^2/\kappa
~.\label{sep}
\end{equation}

A CNOT gate performs a unitary operation in which the value of one qubit is
changed iff the control bit is in state $|1\rangle$. We choose the
first qubit as the control bit which means that the gate should exchange
the
states $|010 \rangle$ and $|011\rangle$, while the states $|000\rangle$ and
$|001\rangle$ remain unchanged. (Here the state $|0j_1j_2 \rangle$
describes a system with no photons in the cavity while the atoms are in
state $|j_1\rangle_1|j_2\rangle_2$).
Eqs.~(\ref{Hcond}), (\ref{Heff}) and (\ref{Hlaser})
and the choice
\begin{equation} \label{freqs}
\Omega^{(1)}_1-\Omega^{(2)}_{1}=\sqrt{2}\, \Omega,~
\Omega^{(2)}_{0}=\sqrt{2} \, \Omega, {~\rm and ~} \Omega^{(1)}_{0}=0
\end{equation}
for the Rabi frequencies lead together with $\Gamma_{\rm cav}=0$ to the
effective Hamiltonian
\begin{eqnarray} \label{Heff3}
H_{\rm eff} &=& {\hbar \over 2} \,
\Big[ \, \Omega \left( |010 \rangle\langle 0\ra| - |0\ra\rangle\langle 011|
\right) +{\rm h.c.} \, \Big] ~.
\end{eqnarray}
A single laser pulse of length $T = \sqrt{2}\pi/|\Omega|$ therefore yields
the desired time evolution operator for the CNOT,
\begin{eqnarray}
U_{\rm eff}(T,0) &=& |010 \rangle\langle 011| + {\rm h.c.} \label{U}
\end{eqnarray}
Here $H_{\rm eff}$ is Hermitian and due to Eq.~(\ref{29}) the probability
for a photon emission during
the laser pulse is, within the approximations made, not possible.

It might be helpful to illustrate the mechanism which confines the dynamics to
the DFS in more
detail.
The time evolution of the system
under the condition of no photon emission is given by the conditional
Hamiltonian $H_{\rm cond} + H_{\rm laser \, I}$. The full equations of 
motion resulting from
Eq.~(\ref{phit}) reveal that
only the amplitudes of DF states change slowly in time, on a time scale
proportional to $1/|\Omega|$. If the system is initially in a DF state the
laser pulse excites the
states outside the DFS. Then the excitation  is transfered
with a rate proportional $g$ into states in which the cavity mode is
populated. Those states are immediately emptied by one of the following
two mechanisms: One possibility is that a photon leaks out through the
cavity mirrors. But, as long as the population of the cavity field is
small, the leakage of a photon through the cavity mirrors is unlikely to
take place. With a much higher probability the excitation of the cavity
field vanishes during the conditional time evolution due to the term $-
{\rm i} \hbar \kappa \, b^\dagger b$ in the conditional Hamiltonian in
Eq.~(\ref{Hcond}). No population can accumulate outside the DFS.

\noindent
\begin{minipage}{3.38truein}
\begin{center}
\begin{figure}
\epsfig{file=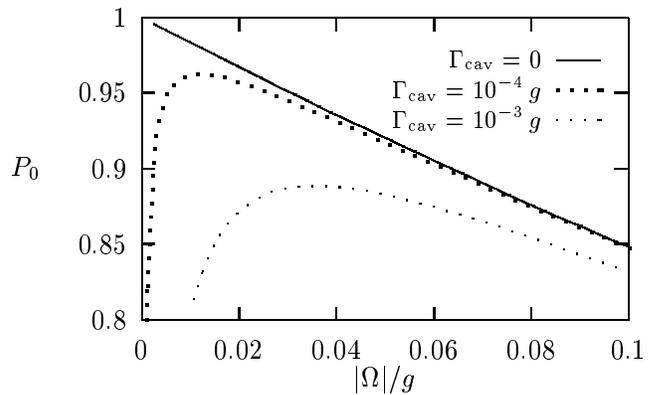,height=5.2cm} \\[0.2cm]
\caption{
The probability for no photon emission during a CNOT operation as a
function of the Rabi frequency $\Omega$ for $\Omega^{(2)}_0=\sqrt{2}\Omega$,
$\Omega^{(1)}_{1}=-\Omega^{(2)}_{1}=\Omega/\sqrt{2}$, $\kappa=g$, and different
values of $\Gamma_{\rm cav}$. The system is initially in the state $|010
\rangle$.}\label{fig.p0}
\end{figure}
\end{center}
\end{minipage}
\vspace*{0.1cm}

We also derived the time evolution of the DF states by
adiabatically eliminating the amplitudes of all non-DF states. This is
possible due to
the frequency scale separation (\ref{sep}). To lowest order in
$\Omega/\kappa$ and $\Omega\kappa/g^2$ we recover Eq.~(\ref{U}). The
more precise result including the next higher order
allows for an optimisation of the gate operations \cite{us99.2}.

If one assumes $\Gamma_{\rm cav} \neq 0$ the state $|0\ra \rangle$ does
not correspond to a DF state and a photon may be emitted during the gate
operation. In addition finite parameters of $g$ and $\kappa$ may lead to
the leakage of photons through the cavity mirrors. These effects have been
taken into account in Fig.~\ref{fig.p0} which results from a numerical
solution of the Schr\"odinger equation (1) and shows the
probability for no photon emission during a single CNOT operation. Here the
initial state $|010\rangle$ was chosen.
The figure confirms  that for
vanishing spontaneous emission the
probability of success becomes arbitrarily close to unity if $|\Omega|$ is made
small. For finite $\Gamma_{\rm cav}$, spontaneous emission is the limiting
factor due to the increasing duration of the
operation for small $\Omega$.
If no photon is emitted during the gate operation, the fidelity of the
state at the end of the pulse
compared to a state expected for an ideal CNOT operation is very high. For
the parameters used in Fig.~\ref{fig.p0} the final amplitude of the desired
state $|011\rangle$ is always higher than $98 \, \%$ \cite{us99.2}.

Finally, we point out that there exists a very simple scheme in which the transition between a DF state and non-DF states is also strongly inhibited, a three-level atom  with a V configuration. One transition between the ground state and a metastable state of the atom is driven by a very weak laser field with Rabi frequency $\Omega_{\rm w}$, while a laser with a very high Rabi frequency $\Omega_{\rm s}$ couples the ground state to a level with a high decay rate $\Gamma_{\rm s}$. In this scheme the metastable state corresponds to the DFS, while $|\Omega_{\rm s}|$ plays the role of the coupling constant $g$ and $\Gamma_{\rm s}$ the role of $\kappa$. Once in a metastable state the atom remains there for a long time proportional to $(|\Omega_{\rm s}|/|\Omega_{\rm w}|)^2$ \cite{almut}.This is known as a {\em macroscopic} dark period and the scheme has been used to test the quantum Zeno effect experimentally. \cite{dark}. Equally, we expect for the scheme proposed here that the mean time before photon emission is proportional to $(g/|\Omega|)^2$ and is much longer than the gate duration which is proportional to $g/|\Omega|$. This is shown explicitly in Ref.~\cite{us99.2} and encourages us to believe that our proposal is experimentally feasible. Due to the correspondence of these schemes we could also describe our proposal as ``quantum computing in a dark period".

Our system also fulfills the remaining criteria
for a quantum computer \cite{Vinc}. The single qubit rotation, which together with the CNOT forms a ``universal'' set of quantum gates, can
be performed with the help of an adiabatic population transfer
\cite{vit} - a technique which has been realised with high accuracy in experiments \cite{bergmann}. It requires two laser pulses and the laser fields couple to the 0-2 and 1-2 transition, respectively, with the same large detuning \cite{us99.2}.
The read out of the information stored in the qubits can be
realised with an electron shelving technique \cite{be}.
The system is scalable, with well characterised qubits, can be prepared in a defined initial state, and the relevant decoherence time is much longer
than the gate operation time if condition (\ref{sep}) can be achieved.
{\em In summary}, we have made a proposal for quantum computing using dissipation. Why the system remains in a DFS can be understood in terms of the quantum Zeno effect.

{\em Acknowledgment:}
We thank F.~Haake, W.~Lange, R.~Laflamme, D.A.~Lidar, M.B.~Plenio,
and T.~Wellens for discussions. Part of this work was done at
the ESF-Newton Institute Conference in Cambridge, partially
supported by the ESF. This work was also supported by the A.~v.~Humboldt Foundation,
by the European Union, by the UK Engineering and Physical Sciences Research
Council, and by the Sonderforschungs\-be\-reich 237 ``Unordnung und
gro{\ss}e Fluktuationen".

\end{multicols}

\end{document}